# Magnetic oscillations Excited by Concurrent Spin Injection from a Tunneling Current and a Spin Hall Current

M. Tarequzzaman[1, 2], T. Böhnert[1], M. Decker[3], J. D. Costa[1], J. Borme[1], B. Lacoste[1], E. Paz[1], A. S. Jenkins[1], S. Serrano-Guisan[1], C. H. Back[3], R. Ferreira[1] and P. P. Freitas[1, 2]

[1]International Iberian Nanotechnology Laboratory (INL), Ave. Mestre Jose Veiga, 4715-330, Braga, Portugal

[2]Physics Department, Instituto Superior Tecnico (IST), Universidade de Lisboa, 1000-029, Lisbon, Portugal

[3]Institut für Experimentelle Physik, Universität Regensburg, D-93040, Regensburg, Germany

## Abstract

In this paper, a 3-terminal spin-transfer torque nano-oscillator (STNO) is studied using the concurrent spin injection of a spin-polarized tunneling current and a spin Hall current exciting the free layer into dynamic regimes beyond what is achieved by each individual mechanism. The pure spin injection is capable of inducing oscillations in the absence of charge currents effectively reducing the critical tunneling current to zero. This reduction of the critical charge currents can improve the endurance of both STNOs and non-volatile magnetic memories (MRAM) devices.

It is shown that the system response can be described in terms of an injected spin current density $J_s$ which results from the contribution of both spin injection mechanisms, with the tunneling current polarization $p$ and the spin Hall angle $\theta$ acting as key parameters determining the efficiency of each injection mechanism. The experimental data exhibits an excellent agreement with this model which can be used to quantitatively predict the critical points ($J_s$ = -2.26±0.09 × $10^9$ $\hbar/e$ A/m$^2$) and the oscillation amplitude as a function of the input currents. In addition, the fitting of the data also allows an independent confirmation of the values estimated for the spin Hall angle and tunneling current polarization as well as the extraction of the damping α = 0.01 and non-linear damping Q = 3.8±0.3 parameters.

# INTRODUCTION

Recent reports demonstrate that the Spin Hall Effect (SHE) can be used to generate pure spin currents, capable of exerting a spin transfer torque (STT) that induces oscillations in a ferromagnetic layer[1,2]. This pure spin current is created by a charge current in a nonmagnetic material with strong spin-orbit coupling where up and down spins are scattered in opposite directions resulting in a spin current orthogonal to the electrical current[2–6]. A central challenge is to quantify the efficiency of the charge current to spin current conversion, which results from the difficulty of measuring spin currents. The spin-orbit material is characterized by a material property called the spin Hall angle, which quantifies the ratio between the generated spin current density ($J_s^{\text{spin Hall}}$) at an applied charge current density ($J_c^{\text{spin Hall}}$). The spin Hall angle is expressed as $\theta = (e/\eta) J_s^{\text{spinHall}} / J_c^{\text{spinHall}}$ with the charge of the electron $e$ and the reduced Plank constant $\hbar$ ensuring dimensional consistency. Several techniques have been used to quantify $\theta$ of transition metals such as Au, Pd, Pt, Ta, and W. A particularly interesting material is Ta since it is a typical cap and seed layer in magnetic tunnel junction (MTJ) devices and in direct contact with the ferromagnetic free layer. The reported $\theta$ values of Ta are in a wide range of $1.4\% < \theta < 15\%$, primarily due to dependences on the crystalline phase[6–9].

From an application point of view, the interest in the SHE is fueled by the possibility of using pure spin currents to excite persistent magnetization oscillations as required to drive ultra-tunable microwave STNOs[1,10,11] or to induce the switching of ferromagnetic layers to write non-volatile magnetic memories (MRAM)[6,12,13]. Such devices are typically manufactured using MTJ nanopillars, using a spin-polarized tunneling current as a source of excitation of the free layer. The onset of dynamic effects can only be observed if the tunneling spin current exerts a torque in the free layer that compensates the Gilbert damping. For in-plane ferromagnetic free layers, this condition can be expressed as[14]:

$$\eta J_{c,\text{crit}}^{\text{tunneling}} \, \hbar/e = 2\mu_0 M_s \, \alpha t \left( M_{\text{eff}}/2 + H_{\text{app}} \right). \qquad \text{(Eq. 1)}$$

Here, $\mu_0$ is the permeability of free space, $\hbar$ is the Planck constant, $M_s$ is the free layer saturation magnetization, $\alpha$ is the Gilbert damping constant, $t$ is the thickness of the free, $H_{\text{app}}$ is the applied field along easy axis, $J_{c,\text{crit}}^{\text{tunneling}}$ is the critical tunneling current density and $\eta$ stands for the spin transfer efficiency associated with the injection of the spin-polarized tunneling current from the reference layer into the free layer. The effective magnetization of free layer is given by $M_{\text{eff}} = (M_s - 2 K_p/(\mu_0 M_s))$ with the perpendicular magnetic anisotropy ($K_p$). This value is determined from saturation field measurements of magnetic films of the same thickness[15].

The injected spin current in the free layer is limited by the irreversible dielectric breakdown of the tunnel barrier. Achieving the onset of dynamic oscillations before the dielectric breakdown of the tunnel barrier requires a large spin injection efficiency $\eta$, which in turn is related to the tunneling current polarization $p$ by $\eta = p/2/(1 + p^2 \cos\varphi)$ with $\varphi = 0°$ in the low resistance configuration and $\varphi = 180°$ in the high resistance configuration[16,17].

MTJ nanopillars fabricated on a Ta micro-stripe adjacent to the free layer in a 3-terminal geometry allows the independent injection of the tunneling current and the spin Hall current in the free layer of the STNOs. In this geometry, the onset of oscillations in the free layer (the left-hand side of Eq. 1) can be expressed by the injected critical spin current density, which is the sum of both contributions:

$$J_s = J_s^{\text{tunneling}} + J_s^{\text{spin Hall}} = \left( \eta \, J_c^{\text{tunneling}} + \theta \, J_c^{\text{spin Hall}} \right) \hbar/e. \qquad \text{(Eq. 2)}$$

Eq. 2 shows that the injection of a spin Hall current lowers the tunneling current required to onset oscillations, a mechanism that can be used to improve the endurance of both STNOs and MRAM cells and improve the output power of STNOs. Furthermore, combining both injection mechanisms can result in injected spin current densities that are beyond those achievable with each isolated mechanism, allowing the exploration of otherwise unreachable dynamic states. For large enough spin Hall currents oscillations can be even achieved in the absence of any tunneling charge current.

The ratio $\theta/\eta$ determines how efficiently the tunneling current can be replaced by a spin Hall current. Typically, the charge to spin conversion efficiency in the available spin-orbit coupling materials is low compared to the spin injection efficiency in MgO based MTJs. As a result, the charge current density in the spin-orbit current line must be considerably larger than the current density tunneling through the tunnel barriers to obtain the similar dynamic states of the free layer. The maximum spin Hall current is typically limited due to the additional heating and stress on the tunnel barrier.

In this work, MTJ nano-pillars with a 200 nm diameter were patterned on a Ta spin Hall micro-stripe (see supplementary information). A stack of 15 Ta/1.4 $Co_{0.4}Fe_{0.4}B_{0.2}$/ MgO /2.2 $Co_{0.4}Fe_{0.4}B_{0.2}$/0.85 Ru/2.0 $Co_{0.7}Fe_{0.3}$/20 $Ir_{0.2}Mn_{0.8}$/5 Ru (thickness in nanometer) were used and incorporated a MgO barrier with a nominal resistance area product ($R \times A$) of 9.7 $\Omega \cdot \mu m^2$. A picture of fabricated device is shown in Fig. 1(a). The devices are measured in a 4-point geometry with magnetic field along the reference layer magnetic direction. The quasi-DC transfer hysteretic curves are consistent with an in-plane magnetization of the 1.4 nm thick $Co_{0.4}Fe_{0.4}B_{0.2}$ free layer. The measured $R{\times}A$ distribution in the final devices is centered around 15 $\Omega\mu m^2$ with the majority of the devices exhibiting a TMR between 100 % and 120 %, as shown in Fig. 1(b). These values are reasonable considering the relatively thin free layer.

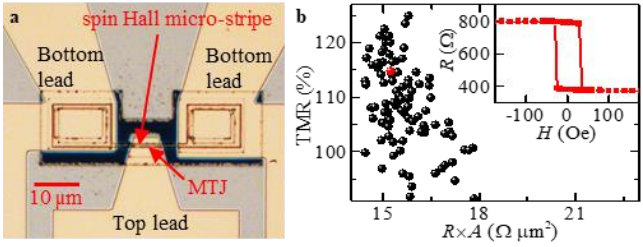

Fig. 1(a) Optical microscope image of the final device. (b) Distribution of TMR and R×A measured in a 4-point geometry of a population of nanopillars with a diameter of 200 nm. The insert shows the 4-point transfer curve of a typical device.

# RESULTS

To describe a concurrent spin injection, according to Eq. 2, both spin injection efficiencies have to be determined. The spin Hall injection efficiency is determined on a 10 Ta/3 $Co_{0.4}Fe_{0.4}B_{0.2}$ (thickness in nanometer) bilayer[18]. The modulation of the effective damping of the ferromagnet[19–21] is measured using the time-resolved magneto-optic Kerr effect (TR-MOKE)[18,22,23]. The spin Hall angle of $\theta = 2.4 \pm 0.14$ % is computed as described in detail in the supplementary information.

The high frequency response of the devices was characterized in a 3-terminal configuration, as shown in Fig. 2(a). The 3-terminal configuration allows an independent control of spin Hall charge current ($I_c^{\text{spin Hall}}$) through the Ta layer and the tunneling charge current ($I_c^{\text{tunneling}}$) through the MTJ nanopillar. In the first step, the frequency response of the free layer was measured by acquiring the power spectral density (PSD) as a function of $I_c^{\text{tunneling}}$, while $I_c^{\text{spin Hall}}$ was set to zero. The $I_c^{\text{tunneling}}$ was swept from 0 mA to -0.7 mA with a magnetic field of -150 Oe biasing the nanopillar in the anti-parallel state. At this current direction, electrons tunnel from the fixed layer to the free layer. The strong free layer magnetization oscillations in the anti-parallel state are in agreement with theoretical predictions[14]. The frequency spectrum shows a peak at ~2.8 GHz with an output power that increases with increasingly negative charge current through the tunnel barrier, as shown in Fig. 2(b) for a particular nanopillar. At the highest probed negative tunneling current, free layer oscillations with a matched output power ($P_{\text{matched}}$) of 19 nW and a linewidth ($\Gamma$) of 100 MHz are observed. To calculate the $P_{\text{matched}}$, the power spectral density (PSD) of the oscillation peaks is integrated and the impedance mismatch between the device and the load is taken into account[24,25]

The output power is sensitive to $I_c^{\text{tunneling}}$ in this device that the usual methods to extract the critical current, such as fitting the normalized inverse power in the thermally activated region[26,27] or fitting $\Gamma$ in the STT excited region[28] lead to inconclusive results. It seems the first method is not feasible here due to very small critical currents and the challenge to obtain enough data points at low tunneling currents for the fit. The critical currents will be discussed later in the manuscript, but one can conclude at this point that tunnel currents of just -0.15 mA are enough to excite oscillations with an integrated power above 1 nW. This corresponds to a tunneling current density of $J_c^{\text{tunneling}} = -4.65 \times 10^9$ A/m². This current density is low to excite oscillations compare to literature[29], but this is not surprising. The thin free layer leads to a low critical current density according to Eq. 1[26], and the MgO barrier thickness is in an intermediate range with a large TMR, which also contributes to decrease the critical current density as reported recently[25].

To estimate $J_s$ using Eq. 1, the $\eta$ has to be determined as a function of $I_c^{\text{tunneling}}$. From the measured low bias TMR value of 102 % and Jullière's model[30] a tunneling spin polarization of $p = 58$ % was estimated. This leads to a spin injection efficiency of $\eta = 44$ % in the anti-parallel configuration, with low bias. However, considering the bias dependence of the TMR, the value of $\eta$ decreases with the absolute value of $I_c^{\text{tunneling}}$ until it reaches 28 % at the maximum current of 0.7mA. As a result a matched power of 1 nW was achieved at $J_s = -1.8 \times 10^9\ \eta/e$ A/m². According to Eq. 2 and the experimental values obtained for $\theta$, a charge current density in the Ta micro-stripe of $J_c^{\text{spin Hall}} = -73 \times 10^9$ A/m² injects an equivalent $J_s$ into the free layer. At this value the spin Hall effect should excite oscillations without any tunneling current. However, a minimum tunneling current of -50 μA is required to transduce the magnetization dynamics of the free layer into high frequency electrical signals, which can be measured in the spectrum analyzer. This reduces the expected oscillation onset to $J_c^{\text{spin Hall}} = -46 \times 10^9$ A/m² (-0.7 mA).

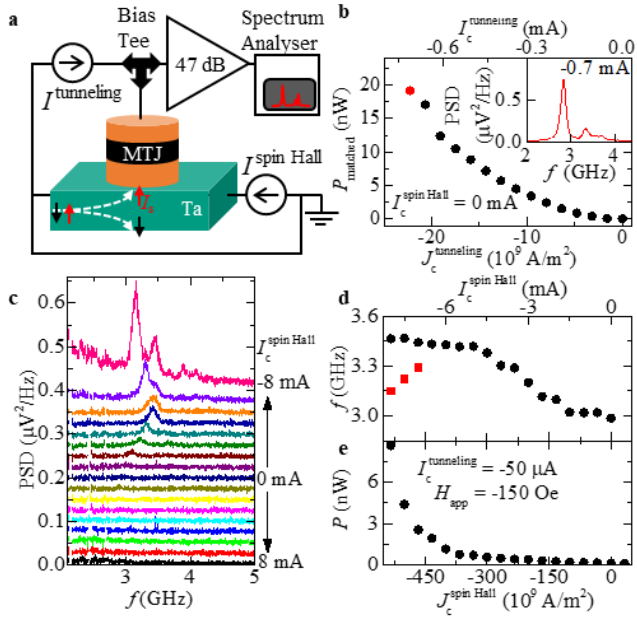

Fig. 2**(a)** Schematic representation of the microwave emission measurement circuit setup for spin-polarized current induced nano-oscillator devices. **(b)** Output power as a function of the $I_c^{tunneling}$ in the absence of a spin Hall current. The inset shows a frequency spectrum obtained at $I_c^{tunneling}$ = -0.7 mA. **(c)** The PSD measured for $I_c^{spin\,Hall}$ ranging from +8 mA to -8 mA at a constant $I_c^{tunneling}$ of -50 µA and $H_{app}$ = -150 Oe. **(d-e)** Frequency $f$ and integrated matched power $P_{matched}$ as a function of $J_c^{spin\,Hall}$. The red and squares indicate the development of a second oscillating mode. Both oscillations modes are fitted to determine the $P_{matched}$.

To confirm this prediction, PSD measurements as a function of $I_c^{spin\,Hall}$ were performed on the same nanopillar in the anti-parallel state. The $I_c^{spin\,Hall}$ is increased with an alternating sign from 0 mA to ±8 mA. Oscillation peaks are only observed for one polarity of the current and they are visible only below $I_c^{spin\,Hall}$ = -2 mA (-133 × 10$^9$ A/m$^2$) as shown in Fig. 2 (c). This excludes merely thermal excitations of the ferromagnetic resonance as discussed in the literature[1]. However, the reference output power of 1 nW is achieved at a value much larger than expected. This apparent discrepancy is due to the fact that the output power depends not only on the tunneling spin current, but also on the tunneling charge current. Therefore, the comparison of the oscillations due to tunneling current and spin Hall current requires more attention than the simple model presented above. This point is discussed in more detail in the following section.

To quantify the obtained results, $P_{matched}$ and $f$ were extracted, as shown in Fig. 2(d-e). The oscillation peak increases with increasing negative $J_c^{spin\,Hall}$ as expected. At $J_c^{spin\,Hall}$ values below -6 mA multiple oscillations modes are detected. This behavior can be an effect related to the $H_{app}$ and geometry of the nanopillars. The two largest peaks are fitted and considered for the $P_{matched}$ calculations. In the steady state STT oscillation, the typical redshift[1] of the frequency as the magnitude of $J_c^{spin\,Hall}$ increases is observed. At large $J_c^{spin\,Hall}$ values $P_{matched}$ reaches 8.6 nW with Γ ≈ 146 MHz. This power is 5-times larger than the values reported in the literature[1,5,11,31] for spin torque oscillators based on SHE (macrospin and nano-constriction oscillators).

The next step is to combine the both spin current source in the same nanopillar. The same external field of -150 Oe was used to set the MTJ nanopillar in the anti-parallel state. The PSD was then acquired while $I_c^{spin\,Hall}$ increased with an alternating sign to ±5 mA in steps of 0.5 mA and $I_c^{tunneling}$ increased with alternating sign ±0.7 mA in steps of 0.05 mA. For each acquired spectrum the output power of the precession peak was integrated and plotted in Fig. 3(a). The results confirm the prediction that the magnitude of microwave oscillators increase with increasingly negative $I_c^{tunneling}$ as well as increasingly negative $I_c^{spin\,Hall}$ with a maximum $P_{matched}$ of 48 nW, which exceeds both maximum $P_{matched}$ obtained with a $I_c^{tunneling}$ tunneling current (19 nW as shown in Fig. 2(b) as well as the maximum $P_{matched}$ obtained as a function of the spin Hall current (8.6 nW as shown in Fig. 2(e)). Note that this larger output power is reached despite each current being limited to lower values than in the previous experiments with each isolated current source. In general, the oscillatory behavior of this sample appeared to be completely reversible. However, in other samples, irreversible barrier deterioration and de-pinning of the fixed layer was observed at larger current values. These are most likely the result of the combined Joule heating of both currents, which could be reduced with further improvements to the spin Hall micro-stripe.

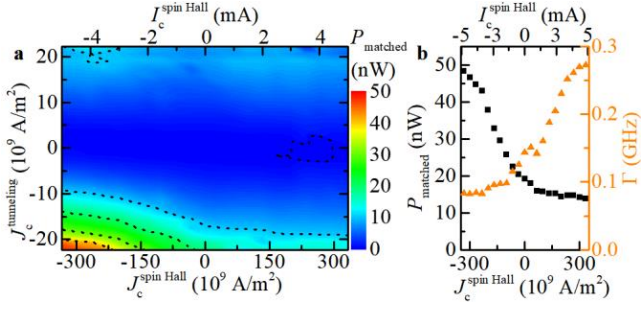

Fig. 3 (a). Color map shows the integrated output power $P_{matched}$ due to the combined excitation by $J_c^{spin\ Hall}$ and $J_c^{tunneling}$. The black dashed lines represent the contour lines of equal power. (b) $P_{matched}$ and linewidth ($\Gamma$) as a function of $J_c^{spin\ Hall}$ at MTJ bias current density $J_c^{tunneling}$ = -22.3 × 10$^9$ A/m$^2$ ($I_c^{tunneling}$ = -0.7 mA) and fixed magnetic field $H_{app}$ = -150 Oe.

Fig. 3(b) displays $P_{matched}$ and $\Gamma$ as a function of $J_c^{spin\ Hall}$ for a fixed value of $J_c^{tunneling}$ = -22.3 × 10$^9$ A/m$^2$. The $J_c^{spin\ Hall}$ can be used to generate a spin current that either reinforces or damp the oscillations depending on the polarity. At negative $J_c^{spin\ Hall}$ the generated spin current reinforces oscillations and reaches $P_{matched}$ of up to 48 nW with a minimum $\Gamma$ of 83 MHz. With increasing $J_c^{spin\ Hall}$ (positive) the damping of the oscillations increases and $P_{matched}$ decreases, as expected from the theoretical models[32,33]. For positive Hall currents $P_{matched}$ reaches an almost constant power of 13.8 nW with $\Gamma$ > 200 MHz. This can be due to the thermal fluctuations of the device reported previously[34–36].

Although the origin of the thermal fluctuations remains unclear, the thermal fluctuations determined at positive spin Hall values and subtracted from $P_{matched}$ to determine the STT dependent contributions ($P_{STT}$) (see supplementary information). The resulting matched rms microwave power $P_{STT}$ can be described by the following equation[37,38]:

$$P_{STT} = I^2 \Delta R^2 / 32R, \quad \quad (Eq.\ 3)$$

with the applied tunneling current $I$ and the static resistance of the device $R$, which is approximately the MTJ resistance in the high resistive state. The observed microwave power results from the resistance oscillation $\Delta R$ due to the magnetization oscillations of the free layer. The factor 32 results from the combination of a factor of 8 due to the peak-to-peak to rms conversion and a factor of 4 due to the power splitting in a matched circuit. The normalized oscillation amplitude can finally be expressed as a ratio between $\Delta R$ and $\Delta R_{max}$ = $R_{AP}$ - $R_P$ as expressed by:

$$\Delta R / \Delta R_{max} = \sqrt{32 R P_{STT}} / I \Delta R_{max}. \quad \quad (Eq.\ 4)$$

Expressing the oscillation in the normalized amplitude $\Delta R/\Delta R_{max}$, rather than $P_{STT}$, allows the direct comparison of all the spin current configurations since this normalization accounts for the remaining tunneling current dependencies. Note that $\Delta R_{max}$ and $\eta$ were determined as a function of $J_c^{tunneling}$ to remove the influence of the decreasing TMR ratio. Any value of $\Delta R/\Delta R_{max}$ describes now a certain excitation amplitude and the contours lines in Fig. 4(a) visualizes point of equal excitation at varying $J_c^{spin\ Hall}$ and $J_c^{tunneling}$. As a consequence of Eq. 2 the slopes of these contour lines correspond to the ratio of $\theta/\eta$. The previously obtained $\theta$ = 2.4 % and the $\eta$ values between 44 % and 28 % (depending on the bias current) are in agreement with this prediction. Thus, the relation given by Eq. 2 seems to be valid and can be used to convert the charge current densities into $J_s$ values. The ratio $\Delta R/\Delta R_{max}$ can be now expressed as a function of $J_s$, as shown in Fig. 4(b).

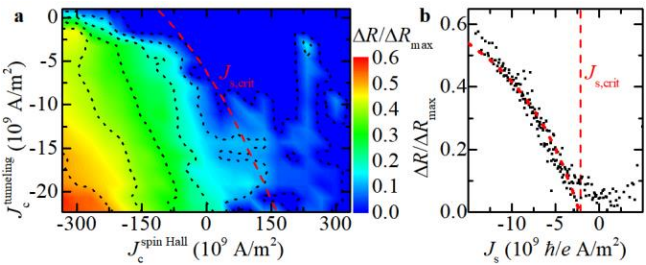

Fig. 4 (a) Color map of the ratio $\Delta R/\Delta R_{max}$. The slope of the contour lines of equal excitation corresponds to the ratio of $\theta/\eta$. The red dashed line follows the combinations of $J_c^{tunneling}$ and $J_c^{spin\ Hall}$ that result in the critical spin current $J_{s,\ crit}$. (b) The ratio $\Delta R/\Delta R_{max}$ versus the total spin current $J_s$ in the MTJ determined by Eq. 2 as described in the manuscript. $J_{s,\ crit}$ is determined from the fit as described in the manuscript.

The consistency of the result clearly shows that both, $J_\text{c}^\text{tunneling}$ and $J_\text{c}^\text{spin Hall}$, contribute to the STT that excites the microwave oscillations, which can be described by Eq. 2. It should be emphasized that only a correct ratio of $\theta/\eta$ leads to the presented relation and any deviations would lead to a significant spread of the observed homogenous behavior in Fig. 4(b). In the proximity of the critical spin current density ($J_\text{s,crit}$) the ratio $\Delta R/\Delta R_\text{max}$ is predicted to be equal to $p_0 = (J_\text{s}-J_\text{s,crit})/(J_\text{s}+J_\text{s,crit}\cdot Q)$, with the non-linear damping constant $Q$[39,40]. This prediction describes the results very well (see Fig. 4(b)) and the fit results in $J_\text{s,crit} = -2.26\pm0.09 \times 10^9$ $\hbar/e$ A/m$^2$ and a non-linear damping $Q = 3.8\pm0.3$ which is comparable to literature[40]. From Eq. 1 a damping constant of 0.008 is determined for a nominal nanopillar diameter of 200 nm. In reality, the nanopillar diameter could be up to 40 nm smaller, due to details of the nanofabrication process, which leads to a damping constant of 0.01. These damping constants are in excellent agreement with the literature[7,41]. The combinations of $J_\text{c}^\text{tunneling}$ and $J_\text{c}^\text{spin Hall}$ that result in spin currents of $J_\text{s,crit}$ are indicated by the curve in Fig. 4(a). The normalized oscillation amplitude $\Delta R/\Delta R_\text{max}$ appears to be a very important step by focusing on the excitation and excluding unrelated effects from the discussion. Although the output power remains a central parameter for applications, this normalized oscillation amplitude seems to be crucial for any quantitative analysis.

# CONCLUSIONS

A 3-terminal MTJ based STNO with a concurrent spin injection from a spin-polarized tunneling current and a spin Hall current was experimentally demonstrated. This combinational mechanism strengthens the total spin current and effectively excites the free layer into dynamic regimes and maximizes total output power. The output power obtained with concurrent spin injection is six-fold higher than the output power achieved with spin Hall currents and two-fold higher than the output power achieved with tunneling currents. A total output power of 48 nW with linewidth of 83 MHz is achieved for the highest (negative) values of $J_\text{c}^\text{tunneling}$ and $J_\text{c}^\text{spin Hall}$. The addition of spin Hall current on the system greatly reduced the critical charge current virtually to zero. Furthermore, the data exhibits an excellent quantitative agreement with theoretical models expressed as a function of the total spin current injected into the free layer. The spin current can be calculated from the input tunneling current and spin Hall current and is used to predict the critical points and oscillation amplitude of the system. The quantitative model confirms the values of key parameters in the system, such as the spin polarization of the tunneling current $p$ and the spin Hall angle $\theta$. From the data fittings, it is possible to extract the critical spin current density, the damping constant and the non-linear damping of the system. All in all, this work is expected to improve the current understanding of STNO devices taking profit of spin Hall currents as well as to contribute to a guided improvement of the STNO properties aiming at practical applications.

# ACKNOWLEDGEMENTS

The research leading to these results has received funding from the European Union Seventh Framework Programme [FP7-People-2012-ITN] under Grant agreement No. 316657 (SpinIcur). M. Tarequzzaman thanks the European Union for funding the European Union Seventh Framework Programme [FP7-People-2012-ITN] and ON2 project INTEGRATION (grant NORTE-07-0124-FEDER-000050).